\documentclass[aps,prb,twocolumn,superscriptaddress,showpacs,amsmath,amssymb]{revtex4}

\usepackage{bm,xspace,graphicx}

\newcommand{\urs}{URu$_2$Si$_2$\xspace}
\newcommand{\tord}{$T_{o}$\xspace}

\newcommand{\tmag}{$T_{m}$\xspace}
\newcommand{\moment}{$\mu_{o}$\xspace}

\newcommand{\mb}{$\mu_{B}/{\rm U}$\xspace}
\newcommand{\dir}[1]{$[ #1 ]$\xspace}
\newcommand{\sigmatoa}{$\sigma$\,$||$\,\dir{100}}
\newcommand{\sigmatob}{$\sigma$\,$||$\,\dir{110}}
\newcommand{\sigmatoc}{$\sigma$\,$||$\,\dir{001}}

\begin{document}

\title{
Competition between hidden order and antiferromagnetism in URu$_2$Si$_2$ under uniaxial stress studied by neutron scattering
}
\author{M.\ Yokoyama}
\email[Electronic Address: ]{makotti@mx.ibaraki.ac.jp}
\affiliation{Faculty of Science, Ibaraki University, Mito 310-8512, Japan}

\author{H.\ Amitsuka}
\author{K.\ Tenya}
\affiliation{Graduate School of Science, Hokkaido University, Sapporo 060-0810, Japan}

\author{K.\ Watanabe}
\author{S.\ Kawarazaki}
\affiliation{Graduate School of Science, Osaka University, Toyonaka 560-0043, Japan}

\author{H.\ Yoshizawa}
\affiliation{Neutron Science Laboratory, Institute for Solid State Physics, The University of Tokyo, Tokai 319-1106, Japan}

\author{J.A.\ Mydosh}
\affiliation{Kamerlingh Onnes Laboratory, Leiden University, P.O.\ Box 9504, 2300 RA Leiden, The Netherlands}
\affiliation{Max Plank Institute for Chemical Physics of Solids, 01187 Dresden, Germany}
\date{\today}

\begin{abstract}
We have performed elastic neutron scattering experiments under uniaxial stress $\sigma$ applied along the tetragonal \dir{100}, \dir{110} and \dir{001} directions for the heavy electron compound \urs . We found that antiferromagnetic (AF) order with large moment is developed with $\sigma$ along the [100] and [110] directions. If the order is assumed to be homogeneous, the staggered ordered moment \moment continuously increases from 0.02 \mb ($\sigma=0$) to 0.22 \mb ($0.25\ {\rm GPa}$). The rate of increase $\partial \mu_o/\partial \sigma$ is $\sim$ 1.0 $\mu_B$/GPa, which is four times larger than that for the hydrostatic pressure ($\partial \mu_o/\partial P \sim$ 0.25 $\mu_B$/GPa). Above 0.25 GPa, \moment shows a tendency to saturate, similar to the hydrostatic pressure behavior. For $\sigma\,||\,[001]$, \moment shows only a slight increase to 0.028 \mb ($\sigma = 0.46\ {\rm GPa}$) with a rate of $\sim$ 0.02 $\mu_B$/GPa, indicating that the development of the AF state highly depends on the direction of $\sigma$. We have also found a clear hysteresis loop in the isothermal $\mu_o(\sigma)$ curve obtained for \sigmatob under the zero-stress-cooled condition at 1.4 K. This strongly suggests that the $\sigma$-induced AF phase is metastable, and separated from the ``hidden order" phase by a first-order phase transition. We discuss these experimental results on the basis of crystalline strain effects and elastic energy calculations, and show that the $c/a$ ratio plays a key role in the competition between these two phases.
\end{abstract}


\maketitle
\section{Introduction}
The nature of the phase transition at $T_o=17.5\ {\rm K}$ in \urs (the ThCr$_2$Si$_2$-type, body-centered tetragonal structure) \cite{rf:Palstra85,rf:Schlabitz86,rf:Maple86} is presently one of the most challenging issues in the heavy-electron physics. The elastic neutron scattering experiments \cite{rf:Broholm87,rf:Mason90,rf:Fak96} indicate that the simple type-I antiferromagnetic (AF) order develops below $\sim$ \tord . However, the obtained staggered moment \moment is extremely small ($\sim 0.02-0.04\ \mu_B/{\rm U}$), and incompatible with the large bulk anomalies such as the specific heat jump at \tord ($\Delta C/T_o\sim 300\ {\rm mJ/K^2\ mol}$). This inconsistency has been puzzling many researchers for almost twenty years, i.e., whether the intrinsic order parameter is the tiny magnetic moment or some unidentified ``hidden" degree of freedom. The key to this issue has been recently obtained from the microscopic studies performed under hydrostatic pressure $P$. We found from the neutron scattering experiments that \moment is strongly enhanced by applying pressure from 0.017 \mb ($P=0$) to 0.25 \mb ($P=1.0\ {\rm GPa}$).\cite{rf:Ami99, rf:Ami2000} In parallel, the $^{29}$Si-NMR study revealed that the system is spatially separated into two differently ordered regions below \tord : one is AF with a large moment and the other is non-magnetic. \cite{rf:Matsuda2001,rf:Matsuda2003} The AF volume fraction is found to increase with $P$, roughly in proportion to $\mu_o^2(P)$, while the magnitude of internal field is almost independent of $P$. This indicates that the observed enhancement of the AF Bragg-peak intensities is attributed to the increase of the AF volume fraction, and not of the local AF moment. Simple extrapolation yields the AF volume fraction at ambient pressure of about 1\%, strongly suggesting that this is the true nature of the tiny magnetic moment. Consequently, the remaining 99\% is considered to be occupied by the ``hidden order", which is responsible for the large bulk anomalies at \tord .

The major purpose of the present study is to investigate how these two types of order correlate with each other. In order to find a relevant parameter, we here examine the effects of lattice distortion. So far various ideas for the hidden order parameters have been proposed, including valence transition, \cite{rf:Barzykin95} uranium dimers, \cite{rf:Kasuya97} unconventional spin density waves, \cite{rf:Ikeda98,rf:Virosztek2002} quadrupolar order \cite{rf:Ami94,rf:Santini94,rf:Takahashi2001,rf:Ohkawa99,rf:Tsuruta2000} and charge current order.\cite{rf:Chandra2001,rf:Chandra2001-2} All of them involve a magnetic instability such that the dipolar order may be replaced with the majority hidden order. This switching is expected to be driven by lattice distortion, since the proposed hidden order parameters are tightly coupled to the lattice system. It is thus interesting to investigate the competition between the two types of order by tuning the crystal distortion. 

A second purpose is to find the relationship between the two ordered states. The $^{29}$Si-NMR results indicate that the AF order develops in parts of the crystal. However, it is not clear whether it is inevitably induced through some coupling with the hidden order parameter, or simply replaced with hidden order by a first order phase transition. In the latter case, hysteretic behavior can be expected in the pressure variations of the AF state. This point, however, was not checked in the previous measurements, \cite{rf:Ami99, rf:Ami2000} where samples were always compressed at room temperature. 

For these purposes, we have performed elastic neutron scattering experiments on \urs , by applying uniaxial stress $\sigma$ under both the stress-cooled and the zero-stress-cooled conditions. We have previously reported some experimental results obtained for weak $\sigma$ up to 0.46 GPa. \cite{rf:Yoko2002,rf:Yoko2003} In the present paper, we have extended the $\sigma$ range up to 0.61 GPa, and also investigated a Rh-doped system U(Ru$_{0.99}$Rh$_{0.01}$)$_2$Si$_2$. The collected results are discussed and interpreted in terms of a lattice distortion (or stress) model involving a distribution of the $c/a$ ratio.

\section{Experimental Procedure}
A single-crystalline sample \urs was grown by the Czochralski pulling method using a tri-arc furnace, and vacuum-annealed at 1000$^{\rm o}$C for a week. Three plates with three different bases of (001), (100) and (110) planes were cut from the crystal by means of spark erosion. The dimensions of the plates are approximately 25 mm$^2$ $\times$ 1 mm. The uniaxial stress $\sigma$ was applied along the \dir{001}, \dir{100} and \dir{110} axes up to 0.61 GPa, by placing the samples between Be-Cu piston cylinders mounted in a clamp-type pressure cell. This cell was used for measuring the temperature variations of the AF state down to 1.5 K under the stress-cooled condition, where the stress was changed at room temperature. 

The elastic neutron scattering experiments were performed by using the triple-axis spectrometer GPTAS (4G) located in the JRR-3M research reactor of Japan Atomic Energy Research Institute. The neutron momentum $k=2.660\ {\rm \AA}^{-1}$ was chosen by using the (002) reflection of pyrolytic graphite (PG) for both monochromating and analyzing the neutron beam. We used the combination of 40'-80'-40'-80' collimators, together with two PG filters to eliminate the higher order reflections. The scans for the stress-cooled process were performed in the ($hk0$), ($h0l)$ and ($hhl$) scattering planes for $\sigma$\,$||$\,\dir{001}, \dir{100} and \dir{110}, respectively. The AF Bragg reflections were obtained by the (100) scans for $\sigma$\,$||$\,\dir{001}, the (100), (102) and (203) scans for $\sigma$\,$||$\,\dir{100}, and  the (111) and (113) scans for $\sigma$\,$||$\,\dir{110}.

For the measurements under the zero-stress-cooled condition, we used a constant-load uniaxial stress apparatus. \cite{rf:Kawa2002} In this apparatus, the Be-Cu pistons in the pressure cell, which is attached to the bottom of the $^4$He cryostat insert, is compressed by an oil-pressure device mounted on the top of the insert via a movable rod made of stainless-steel and tungsten carbide. The load is precisely stabilized by controlling the oil pressure during the measurements. We first cooled the sample down to 1.4 K without compression, and then applied the uniaxial stress along the [110] direction up to 0.4 GPa, keeping the sample at the same temperature. The scans for the zero-stress-cooled condition were performed in the $(hhl)$ scattering plane. The AF Bragg reflections were obtained by the longitudinal scans at the (111) position. The experiments under the zero-stress-cooled condition ($\sigma$\,$||$\,\dir{100}) were also performed on the Rh-doped alloy U(Ru$_{0.99}$Rh$_{0.01}$)$_2$Si$_2$, which was prepared in the same procedure as the pure compound. The (100) magnetic Bragg reflections were investigated by using longitudinal scans in the ($h0l$) scattering plane at 1.4 K.

\section{Experimental Results}
\subsection{Elastic neutron scattering under stress-cooled condition }
\begin{figure}[tbp]
\begin{center}
\includegraphics[keepaspectratio,width=0.4\textwidth]{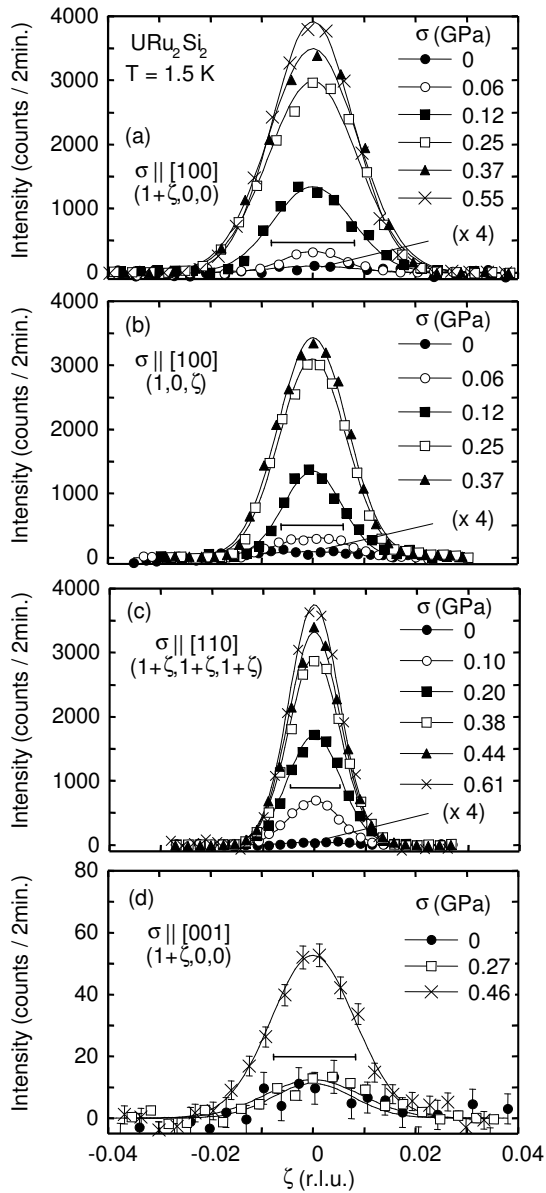}
\end{center}
  \caption{The uniaxial-stress variations of the magnetic Bragg peaks of \urs , obtained from (a) the longitudinal and (b) the transverse scans at the (100) position for $\sigma$\,$||$\,\dir{100}, and the longitudinal scans at (c) the (111) position for $\sigma$\,$||$\,\dir{110} and (d) (100) for $\sigma$\,$||$\,\dir{001} at 1.5 K. The horizontal bars indicate the widths (FWHM) of the resolution limit estimated from the higher-order nuclear reflections. Note that the data for $\sigma=0$ are 4 times enlarged in (a), (b) and (c).}
  \label{fig:profile-sc-uniaxial}
\end{figure}

Figure \ref{fig:profile-sc-uniaxial} shows the $\sigma$ variations of the longitudinal and transverse scans at 1.5 K through the (100) magnetic peak for $\sigma$\,$||$\,\dir{100} and $||$\,\dir{001}, and the longitudinal scans through the (111) peak for $\sigma$\,$||$\,\dir{110}. The instrumental background and the contamination of the higher-order nuclear reflections were carefully subtracted by using the data taken at 40 K. As stress is applied along the \dir{100} direction, the (100) peak intensity markedly increases (Fig.\ \ref{fig:profile-sc-uniaxial}(a) and (b)). The (102) and (203) peaks also develop rapidly (not shown). The intensities of these three magnetic reflections divided by the polarization factor roughly follow the $|Q|$ dependence of the magnetic form factor \cite{rf:Frazer65} of U$^{4+}$. On the other hand, no reflection is observed at (001) position and also in the scans along the principal axes in the first Brillouin zone: $(1+\zeta, 0,0)$, $(1+\zeta,0,1-\zeta)$ and $(2,0,\zeta)$ for $0\le \zeta \le 1$. These results indicate that the type-I AF structure with moments polarized along the $c$ axis is unchanged by the application of $\sigma$\,$||$\,\dir{100}. The development of the magnetic scattering is also observed for $\sigma$\,$\parallel$\,$[110]$ (Fig.\ \ref{fig:profile-sc-uniaxial}(c)). From the same analyses, we confirm that the AF structure is unchanged also for $\sigma$\,$||$\,\dir{110}. In contrast to the compression along the basal plane, the increase of magnetic reflections for $\sigma$\,$\parallel$\,$[001]$ is very small (Fig.\ \ref{fig:profile-sc-uniaxial}(d)), indicating that the AF state strongly depends on the direction of $\sigma$.

The magnetic Bragg peaks observed at (100) and (111) were fitted by the Lorentzian function convoluted with the Gaussian resolution function, to estimate the correlation lengths $\xi$ of the AF moment. The instrumental resolutions are estimated from the widths (FWHM) of higher-order nuclear reflections measured at the corresponding $Q$ positions without PG filters. At ambient pressure, $\xi$ along the \dir{100}, \dir{001} and \dir{111} directions are estimated to be about 150 ${\rm \AA}$, 260 ${\rm \AA}$ and 330 ${\rm \AA}$, respectively. They increase rapidly by applying $\sigma$ along the \dir{100} and \dir{110} directions. Above 0.3 GPa, the peak widths approach the resolution limit ($\sim 1000\ {\rm \AA}$), and the simple fits give the $\xi$ values of approximately 2.5 times larger than those for $\sigma=0$. On the other hand, $\xi$\,$\parallel$\,$[100]$ for $\sigma$\,$\parallel$\,\dir{001} remains around a small value of $\sim$ 230 ${\rm \AA}$ even at 0.46 GPa. These results indicate that the increase of $\xi$ is accompanied by the enhancement of the AF Bragg-peak intensities.
\begin{figure}[tbp]
\begin{center}
\includegraphics[keepaspectratio,width=0.4\textwidth]{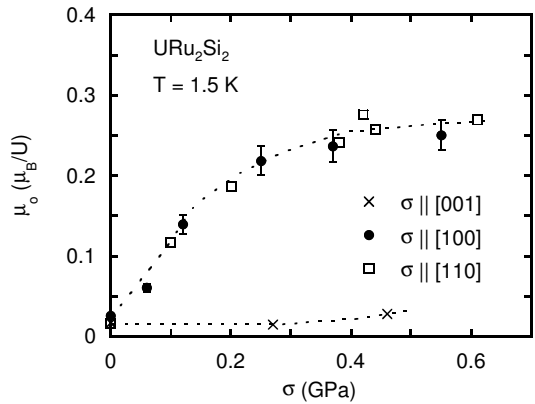}
\end{center}
  \caption{Uniaxial-stress dependence of the staggered moment \moment at 1.5 K. The values of \moment are estimated by assuming homogeneous AF order. The broken lines are guides to the eye.}
  \label{fig:moment-sc-uniaxial}
\end{figure}

Displayed in Fig.\ \ref{fig:moment-sc-uniaxial} is the $\sigma$ dependence of the  staggered moment \moment at 1.5 K. The magnitudes of \moment are obtained from the integrated intensities of the magnetic Bragg peaks at (100) for $\sigma$\,${||}$\,\dir{001} and \dir{100}, and at (111) for $\sigma$\,${||}$\,\dir{110}, which are normalized by the intensities of the weak nuclear (110) reflection for $\sigma$\,${||}$\,\dir{001} and \dir{110}, and (101) for $\sigma$\,${||}$\,\dir{100}. We should note that the \moment values estimated here are based on the assumption of homogeneous AF order. At $\sigma=0$, \moment is 0.020(4) \mb , which roughly corresponds with the values of previous investigations. \cite{rf:Broholm87,rf:Mason90,rf:Fak96} As stress is applied along the \dir{100} direction up to 0.25 GPa, \moment is strongly enhanced to 0.22(2) \mb , and then shows a tendency to saturate above 0.25 GPa. The \moment value at 0.55 GPa is estimated to be 0.25(2) \mb . The $\mu_o(\sigma$) curve for $\sigma\,||$\,\dir{100} is quite similar to that for the hydrostatic pressure. \cite{rf:Ami99,rf:Ami2000} This similarity strongly suggests that the enhancement of \moment under $\sigma$ is also attributed to the increase of the AF volume fraction. However, the estimated rate of increase, $\partial \mu_o/\partial \sigma$ $\sim$ 1.0 $\mu_B/{\rm GPa}$, is much larger than that for the hydrostatic pressure, $\partial \mu_o/\partial P$ $\sim$ 0.25 $\mu_B/{\rm GPa}$. Interestingly, \moment also develops with $\sigma$\,$\parallel$\,$[110]$, tracing the curve for $\sigma$\,$||$\,\dir{100} within the experimental accuracy. For $\sigma$\,$||$\,\dir{001}, on the other hand, \moment slightly increases to 0.028(3) \mb at 0.46 GPa, with a small rate $\partial \mu_o/\partial \sigma$ $\sim$ 0.02 $\mu_B/{\rm GPa}$.

\begin{figure}[tbp]
\begin{center}
\includegraphics[keepaspectratio,width=0.4\textwidth]{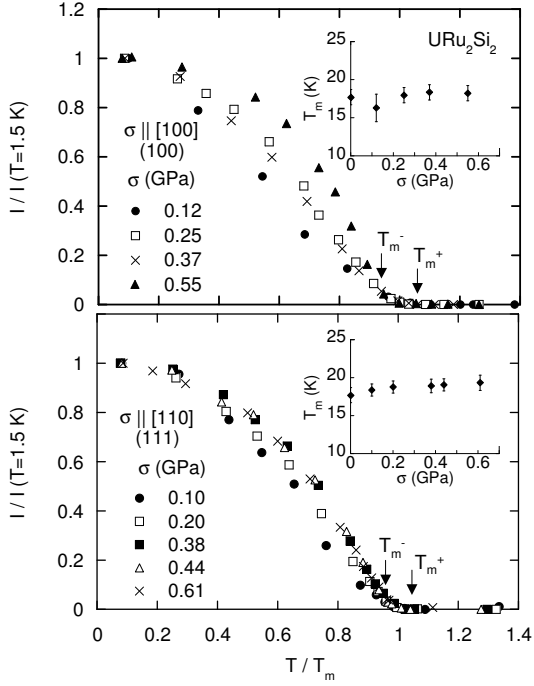}
\end{center}
  \caption{Temperature dependence of the normalized Bragg peak intensities $I/I(1.5\ {\rm K})$ for $\sigma$\,${||}$\,\dir{100} (top) and $\sigma$\,${||}$\,\dir{110} (bottom). Temperature is scaled by $T_m$ defined in the text. The insets show the $\sigma$ dependence of \tmag .}
  \label{fig:temp-uniaxial}
\end{figure}

In Fig.\ \ref{fig:temp-uniaxial}, we plot the normalized Bragg-peak intensity $I/I({\rm 1.5\ K})$ for $\sigma$\,$||$\,\dir{100} and \dir{110} as a function of normalized temperature $T/T_m$, where $T_m$ is defined as the onset temperature of $I(T)$ as follows. Upon cooling, $I(T)$ starts increasing at a temperature $T_m^+$ and exhibits a $T$-linear dependence below $T_m^-\ (<T_m^+)$. The width, $\delta T_m=T_m^+-T_m^-$, of this ``tail" of $I(T)$ is estimated to be 2--3 K, and we define \tmag as the midpoint of $T_m^+$ and $T_m^-$. Although the experimental errors are somewhat large, the $\sigma$ variations of \tmag fall in the range of $\sim\pm1.5\ {\rm K}$ from $T_m(\sigma=0)$ $\sim$ 17.7 K, thereby showing a remarkable contrast with the large $\sigma$ variations of \moment . The observed weak variations of \tmag are not inconsistent with the $\sigma$ variation of \tord ($dT_o/d\sigma=1.26\ {\rm K/GPa}$), which is obtained from the electrical resistivity measurements for $\sigma$\,$||$\,\dir{100}.\cite{rf:Bakker92} For a weak stress range $\sigma \le 0.12\ {\rm GPa}$, the $I(T)$ curves for both the $\sigma$ directions exhibit unusually slow saturation with decreasing temperature. For further compression, $I(T)$ shows a sharper onset and more rapid saturation, pronounced in a rounding curvature for $T/T_m\le 0.6$.

\subsection{Elastic neutron scattering under zero-stress-cooled condition}
\begin{figure}[tbp]
\begin{center}
\includegraphics[keepaspectratio,width=0.4\textwidth]{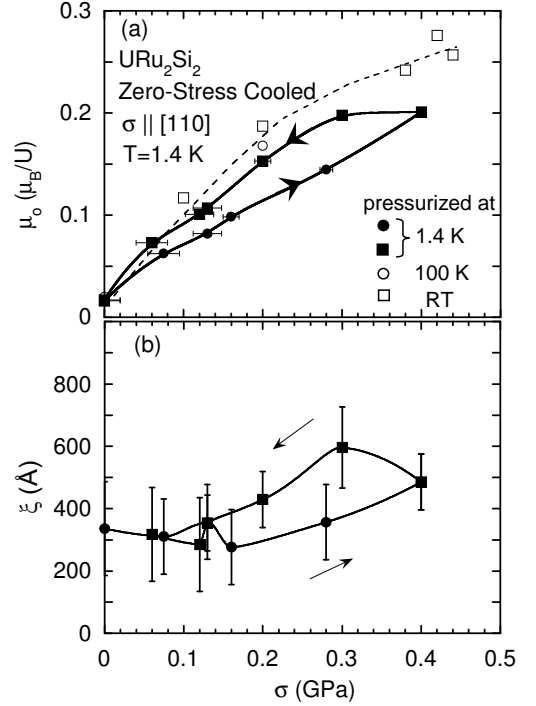}
\end{center}
\caption{The uniaxial-stress variations of (a) the staggered moment \moment and (b) the correlation length $\xi$ along the [111] direction for \sigmatob in \urs , measured at 1.4 K after cooling the sample at $\sigma=0$. The \moment data taken under the stress-cooled condition (pressurized at 100 K and room temperature) are also plotted. The inset of (a) shows the $\sigma$ variations of \moment for \sigmatoa in U(Ru$_{0.99}$Rh$_{0.01}$)$_2$Si$_2$, measured at 1.4 K under zero-stress-cooled condition. The lines are guides to the eye.}
\label{fig:moment-zsc-uniaxial}
\end{figure}

Figure \ref{fig:moment-zsc-uniaxial} (a) shows the $\sigma$ variations of \moment for \sigmatob , obtained from the increasing and decreasing $\sigma$ sweeps at 1.4 K under the {\it zero-stress-cooled} condition. The \moment value is estimated from the integrated intensity of the (111) magnetic Bragg peaks normalized by using the nuclear (110) reflection. The $\mu_o(\sigma)$ curve shows a clear hysteresis loop. As $\sigma$ is applied, \moment develops linearly from 0.016(4) \mb ($\sigma=0$) to 0.20(1) \mb ($\sigma=0.4\ {\rm GPa}$). Upon decompression, on the other hand, \moment shows nearly $\sigma$-independent behavior between 0.4 and 0.3 GPa, and then starts decreasing. After the cycle of compression, \moment returns approximately to the initial value at ambient pressure. The $\mu_o(\sigma)$ curve for the $\sigma$-decreasing process is very similar to that obtained under the stress cooled condition. 

In general, the application of $\sigma$ may increase the crystalline mosaic, which weakens the extinction of reflection and leads to a significant error in the estimation of the intrinsic neutron scattering intensity. Within the pressure range of the present measurements, the intensity of the magnetic (111) peak is always smaller than that of the nuclear (110) reference peak. Normally, the stronger the reflection, the larger the influence of extinction. We, however, observed that the integrated intensity of the (110) peak increases by $15\%$, which is much smaller than that of the (111) peak. In addition, the difference of the (110) peak intensity between the increasing and decreasing $\sigma$ sweeps is within the range of $4\%$, which is also much smaller than that of the (111) peak intensity. The observed enhancement of the magnetic (111) reflection is thus not due to the variation of the extinction effects. We also checked the instrumental error of $\sigma$ between the stress-increasing and decreasing processes by using a strong (220) nuclear reflection. The $\sigma$ variations of the integrated intensity due to the extinction effects show no significant hysteresis, and we confirm that the error of $\sigma$ between the two processes is at most $\pm 0.02\ {\rm GPa}$, as indicated by error bars in Fig.\ \ref{fig:moment-zsc-uniaxial}.

The widths (FWHM) of the (111) magnetic Bragg-peaks are slightly larger than the instrumental resolution, and in Fig.\ \ref{fig:moment-zsc-uniaxial} (b) we show the correlation length $\xi$ of the AF moment along the [111] direction as a function of $\sigma$. At ambient pressure, $\xi$ is estimated to be about 340 ${\rm \AA}$. As $\sigma$ is applied to 0.4 GPa, $\xi$ increases to $\sim 500\ {\rm \AA}$. Upon decompression, it continues to increase, reaches a maximum at $\sim$ 0.3 GPa, and then returns to near the initial value. Although the experimental errors are large, one can see a qualitative correspondence between the $\mu_o(\sigma)$ and the $\xi(\sigma)$ curves. The hysteresis loops observed in the $\mu_o(\sigma)$ and $\xi(\sigma)$ curves strongly suggest that the $\sigma$-induced AF order is metastable and separated from hidden order by a first-order phase transition.

\section{Discussion}
\subsection{Crystal strains under hydrostatic pressure and uniaxial stress}
It is important to remark that the uniaxial stress applied in the $a-a$ plane brings about similar characteristics of the AF order, magnitude of $\mu_o$ as well as its $T$ and $\sigma$ dependences, to those given by hydrostatic pressure. \cite{rf:Ami99,rf:Ami2000} This implies that there is an implicit and common parameter leading to an equivalent effect in the different types of compression. In this subsection, we discuss the crystal strains caused by $P$ and $\sigma$, and propose that the $c/a$ ratio plays an important role in the competition between the two ordered phases.

Within the linear approximation, the uniaxial stresses in the tetragonal crystal symmetry are coupled with the strains by the relation, 
\begin{eqnarray}
\left(
\begin{array}{c}
\sigma_{xx} \\
\sigma_{yy} \\
\sigma_{zz} \\
\sigma_{yz} \\
\sigma_{zx} \\
\sigma_{xy}  
\end{array}
\right)
=
\left(
\begin{array}{cccccc}
  c_{11}&  c_{12}&  c_{13}& & & \\
  c_{12}&  c_{11}&  c_{13}& & & \\
  c_{13}&  c_{13}&  c_{33}& & & \\
  & & & c_{44}& & \\
  & & & & c_{44}& \\
  & & & & & c_{66}\\
\end{array}
\right)
\left(
\begin{array}{c}
\varepsilon_{xx} \\
\varepsilon_{yy} \\
\varepsilon_{zz} \\
\varepsilon_{yz} \\
\varepsilon_{zx} \\
\varepsilon_{xy}  
\end{array}
\right), \label{eq:matrix-elastic}
\end{eqnarray}
where the $\sigma_i$'s, $c_{ij}$'s and $\varepsilon_{j}$'s indicate the uniaxial stresses, elastic constants and strains. The elastic energy symmetrized in the tetragonal point group can be expressed by the form: \cite{rf:Morin88}
\begin{eqnarray}
E_{el} &=&\frac{1}{2}c^{\alpha 1}(\varepsilon^{\alpha 1})^2+c^{\alpha 12}\varepsilon^{\alpha 1}\varepsilon^{\alpha 2}+\frac{1}{2}c^{\alpha 2}(\varepsilon^{\alpha 2})^2\\ \nonumber
&&{}+\frac{1}{2}c^{\gamma}(\varepsilon^{\gamma})^2+\frac{1}{2}c^{\delta}(\varepsilon^{\delta})^2+\frac{1}{2}c^{\epsilon}\{(\varepsilon^{\epsilon}_1)^2+(\varepsilon^{\epsilon}_2)^2\}.
\end{eqnarray}
The definition of the $c^i$'s and $\varepsilon^i$'s are given in Table \ref{tbl:strain}. These notations for the strains are useful in discussing the symmetry of lattice distortion. For example, the strains of $\varepsilon^{\gamma}$, $\varepsilon^{\delta}$, and $\varepsilon^{\epsilon}$ types break the tetragonal symmetry, while the strains of $\varepsilon^{\alpha}$ type change the volume and the $c/a$ ratio, conserving the tetragonal symmetry. We also show in Table \ref{tbl:strain2} the symmetrized strains divided by the stresses, $\varepsilon^{i}/X$ for $X=P$, $\sigma$\,$\parallel$\,$[100]$, $\sigma$\,$\parallel$\,$[110]$ and $\sigma$\,$\parallel$\,$[001]$, calculated from Eq.\ (\ref{eq:matrix-elastic}).
\begin{table}[htbp]
\begin{center}
\begin{tabular}{cc}
\hline \hline
 Strains & Elastic Constants \\
\hline
 $\varepsilon^{\alpha 1}=\frac{1}{\sqrt{3}}(\varepsilon_{xx}+\varepsilon_{yy}+\varepsilon_{zz})$ & $c^{\alpha 1}=\frac{1}{3}(2c_{11}+2c_{12}+4c_{13}+c_{33})$ \\
 $\varepsilon^{\alpha 2}=\sqrt{\frac{2}{3}}(\varepsilon_{zz}-\frac{\varepsilon_{xx}+\varepsilon_{yy}}{2})$  & $c^{\alpha 12}=-\frac{\sqrt{2}}{3}(c_{11}+c_{12}-c_{13}-c_{33})$ \\
 $\varepsilon^{\gamma}=\frac{1}{\sqrt{2}}(\varepsilon_{xx}-\varepsilon_{yy})$ & $c^{\alpha 2}=\frac{1}{3}(c_{11}+c_{12}-4c_{13}+2c_{33})$ \\
 $\varepsilon^{\delta}=\sqrt{2}\varepsilon_{xy}$ & $c^{\gamma}=c_{11}-c_{12}$ \\
 $\varepsilon^{\epsilon}_1=\sqrt{2}\varepsilon_{zx}$  & $c^{\delta}=2c_{66}$ \\
 $\varepsilon^{\epsilon}_2=\sqrt{2}\varepsilon_{yz}$ & $c^{\epsilon}=2c_{44}$ \\
 \hline \hline
\end{tabular}
\end{center}
\caption{The symmetrized strains and elastic constants in the tetragonal symmetry \cite{rf:Morin88}.}
\label{tbl:strain}
\end{table}

Let us now consider the influence of the symmetry-breaking strains $\varepsilon^{\gamma}$, $\varepsilon^{\delta}$, $\varepsilon^{\epsilon}_1$ and $\varepsilon^{\epsilon}_2$ on the AF order. It is obvious that none of them can be caused by $P$ and $\sigma$\,$\parallel$\,$[001]$. On the other hand, \sigmatoa and \sigmatob give rise to $\varepsilon^{\gamma}$ and $\varepsilon^{\delta}$, respectively. Therefore, if the AF order is induced by the symmetry-breaking strains, it should occur only by \sigmatoa and {\sigmatob}, and it is not necessary for their effects to be the same. This is inconsistent with our experimental results: \moment is induced by both $P$ and $\sigma$ (in-plane), and \sigmatoa and \sigmatob have the same effects within the experimental accuracy. We thus conclude that the symmetry-breaking strains are irrelevant to the evolution of the AF phase, at least, in the weak pressure range.

\begin{figure}[h]
\begin{center}
\includegraphics[keepaspectratio,width=0.4\textwidth] {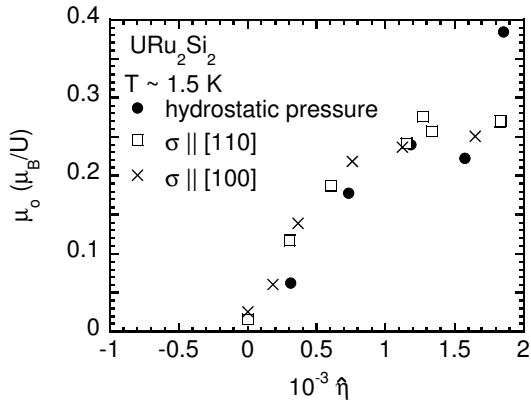}
\end{center}
\caption{The spatially averaged AF moment \moment obtained from the elastic neutron scattering under hydrostatic pressure $P$ and the uniaxial stresses \sigmatoa and [110], plotted as a function of $\hat{\eta}\equiv (\eta-\eta_0)/\eta_0=\varepsilon_{zz}-(1/2)(\varepsilon_{xx}+\varepsilon_{yy})$. }
\label{fig:scale}
\end{figure}

We next consider the variations of the symmetry-invariant strains, $\varepsilon^{\alpha 1}$ and $\varepsilon^{\alpha 2}$, which can be expressed by the relative variations of the unit cell volume, $V$, and the $c/a$ ratio, $\eta$, as follows:
\begin{eqnarray}
\hat{v} &\equiv&\frac{V-V_0}{V_0}=\varepsilon_{xx}+\varepsilon_{yy}+\varepsilon_{zz}=\sqrt{3}\varepsilon^{\alpha 1},\\
\hat{\eta}
&\equiv&\frac{\eta-\eta_0}{\eta_0}=\varepsilon_{zz}-\frac{\varepsilon_{xx}+\varepsilon_{yy}}{2}=\sqrt{\frac{3}{2}}\varepsilon^{\alpha 2},
\end{eqnarray}
where $V_0$ and $\eta_0$ denote the values at ambient pressure. Using the known $c_{ij}$ values of \urs (Table \ref{tbl:elastic}),\cite{rf:Wolf94} we calculated the rates of change in the volume, $\partial \hat{v}/\partial X$, and the $c/a$ ratio, $\partial \hat{\eta}/\partial X$, in Table \ref{tbl:rate}. The calculations show that $\hat{\eta}$ is increased by \sigmatoa and [110] at the same rate, $\partial \hat{\eta}/\partial \sigma \sim 3.0 \times 10^{-3}\ {\rm GPa}^{-1}$. Interestingly, $\hat{\eta}$ is also expected to increase under hydrostatic pressure, because of the Poisson's effect. From the calculations we obtained the relation between the increasing rates: $\partial \hat{\eta}/\partial \sigma \sim 3 \times \partial \hat{\eta} /\partial P$. These features seem to be consistent with the experimental results that \moment are equally enhanced by \sigmatoa and [110], and the relation $\partial \mu_o/\partial \sigma \sim 4 \times \partial \mu_o/\partial P$ holds. The observed $\mu_o(P)$ and $\mu_o(\sigma)$ curves are well scaled by $\hat{\eta}$ (Fig.\ \ref{fig:scale}), indicating that the $c/a$ ratio is relevant to the competition between the two types of order. On the other hand, the volume contraction $\hat{v}$ is irrelevant, because $P$ should exert a stronger influence than $\sigma$, which is inconsistent with the observation. In this context, however, \moment is expected to be suppressed by applying {\sigmatoc}, whereas actually it is almost independent of the stress (Fig.\ \ref{fig:moment-sc-uniaxial}). This can be understood, if the AF phase observed at ambient pressure is caused by irremovable local distortions which are ``pinned" near impurities and defects.
\begin{table*}[htbp]
\begin{center}
\begin{tabular}{c|cccc}
\hline \hline
 $X$ & $P$ & \sigmatoa & \sigmatob & \sigmatoc \\
\hline
 Strain & $(-P,-P,-P,0,0,0)$ & $(-\sigma,0,0,0,0,0)$ & $(-\sigma /2,-\sigma /2,0,0,0,-\sigma /2)$ & $(0,0,-\sigma,0,0,0)$ \\ 
$\varepsilon^{\alpha 1}/X$ &
$-\frac{1}{\sqrt{3}}\frac{c_{11}+c_{12}-4c_{13}+2c_{33}}{-2c_{13}^2+(c_{11}+c_{12})c_{33}}$ & $-\frac{1}{\sqrt{3}}\frac{-c_{13}+c_{33}}{-2c_{13}^2+(c_{11}+c_{12})c_{33}}$ & 
$-\frac{1}{\sqrt{3}}\frac{-c_{13}+c_{33}}{-2c_{13}^2+(c_{11}+c_{12})c_{33}}$ & $-\frac{1}{\sqrt{3}}\frac{c_{11}+c_{12}-2c_{13}}{-2c_{13}^2+(c_{11}+c_{12})c_{33}}$ \\
$\varepsilon^{\alpha 2}/X$ & $-\sqrt{\frac{2}{3}}\frac{c_{11}+c_{12}-c_{13}-c_{33}}{-2c_{13}^2+(c_{11}+c_{12})c_{33}}$ & $\frac{1}{\sqrt{6}}\frac{2c_{13}+c_{33}}{-2c_{13}^2+(c_{11}+c_{12})c_{33}}$ & 
$\frac{1}{\sqrt{6}}\frac{2c_{13}+c_{33}}{-2c_{13}^2+(c_{11}+c_{12})c_{33}}$ & $-\sqrt{\frac{2}{3}}\frac{c_{11}+c_{12}+c_{13}}{-2c_{13}^2+(c_{11}+c_{12})c_{33}}$ \\
 $\varepsilon^{\gamma}/X$ & 0 & $- \frac{1}{\sqrt{2}}\frac{1}{c_{11}-c_{12}}$ & 0 & 0\\
 $\varepsilon^{\delta}/X$ & 0 & 0 & $-\frac{1}{\sqrt{2}}\frac{1}{c_{66}}$ & 0 \\
 $\varepsilon^{\epsilon}_1/X, \varepsilon^{\epsilon}_2/X$ & 0 & 0 & 0 & 0 \\
 \hline \hline
\end{tabular}
\end{center}
\caption{The symmetrized strains divided by stresses induced by hydrostatic pressure and uniaxial stresses in the tetragonal symmetry.}
\label{tbl:strain2}
\end{table*}
\begin{table}[htbp]
\begin{center}
\begin{tabular}{cccccc}
\hline \hline
$c_{11}$ & $c_{33}$ & $c_{44}$ & $c_{66}$ & $c_{12}$ & $c_{13}$ \\
\hline
25.5 & 31.3 & 13.3 & 18.8 & 4.8 & (8.6) \\
\multicolumn{6}{r}{($\times 10^{11}$ erg/cm$^3$)}\\
\hline \hline
\end{tabular}
\end{center}
\caption{The elastic constants at low-temperatures for \urs obtained by the ultrasonic-sound-velocity measurements. \cite{rf:Wolf94} The value for $c_{13}$ was estimated from a comparison between \urs and the isostructural compounds CeCu$_2$Si$_2$ and CeRu$_2$Si$_2$.}
\label{tbl:elastic}
\end{table}
\begin{table}[htbp]
\begin{center}
\begin{tabular}{c|cccc}
\hline \hline
 $X$ &$P$ & \sigmatoa & \sigmatob  &  \sigmatoc \\
\hline
$\partial \hat{v}/\partial X$ & $-7.3$ & $-2.8$ & $-2.8$ & $-1.6$ \\
$\partial \hat{\eta}/\partial X$  & 1.2 & 3.0 & 3.0 & $-4.9$ \\
&\multicolumn{4}{r}{($\times 10^{-3}$ GPa$^{-1}$)}\\
\hline \hline
\end{tabular}
\end{center}\caption{The increasing rate of the symmetry invariant strains induced by hydrostatic pressure and uniaxial stresses, calculated from the elastic constants of \urs .}
\label{tbl:rate}
\end{table}

The magneto-elastic energy $E_{me}$ for the type-I AF order in the tetragonal crystal is given by
\begin{equation}
E_{me}=-D_v\hat{v} M^2-D_\eta\hat{\eta} M^2,
\end{equation}
where $M$ denotes the staggered magnetization and $D_{v, \eta}$ magneto-elastic coupling constants. \cite{rf:Yuasa94} The above consideration implies that $|D_\eta|$ is larger than $|D_v|$ in \urs . This is supported by recent thermal-expansion measurements performed under $P$, which revealed that the $c/a$ ratio significantly increases as the AF phase develops with decreasing temperature.\cite{rf:Motoyama2002,rf:Motoyama2003}

\begin{figure}[tbp]
\begin{center}
\includegraphics[keepaspectratio,width=0.4\textwidth] {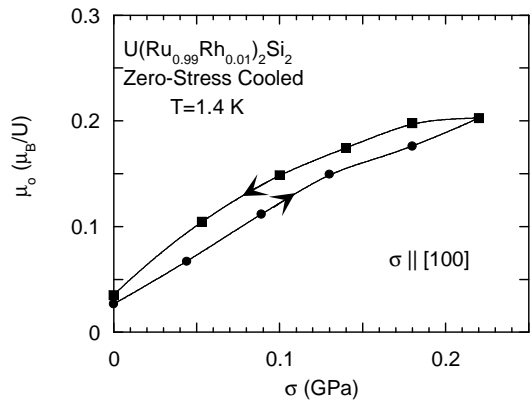}
\end{center}
\caption{The uniaxial-stress ({\sigmatoa}) variations of the staggered moment \moment for U(Ru$_{0.99}$Rh$_{0.01}$)$_2$Si$_2$, measured at 1.4 K after cooling the sample at $\sigma=0$.}
\label{fig:rh}
\end{figure}

The significance of the $c/a$ ratio is also recognized from the behavior of the alloy system U(Ru$_{1-x}$Rh$_x$)$_2$Si$_2$. In this system, the $c/a$ ratio is known to increase as $x$ increases. \cite{rf:Burlet92} For $x\sim 0.02$, $\hat{\eta}$ reaches $\sim 1\times 10^{-3}$: the value at which the AF phase is fully induced in the pure compound (see Fig.\ \ref{fig:scale}). Correspondingly, the AF phase is found to develop at $x \sim 0.015$. \cite{rf:Yoko2004} To test the relevance of the ``chemical stress" to the phenomena, we applied uniaxial stress ($||$\, \dir{100}) to the alloy U(Ru$_{0.99}$Rh$_{0.01}$)$_2$Si$_2$. We observed that $\mu_o(T = 1.4\ {\rm K})$ steeply increases with $\sigma$, from 0.026(3) to 0.20(2) \mb , and the saturation of \moment is more abrupt than that for the pure system (Fig.\ \ref{fig:rh}). These facts indicate that the axial strain, which is generated by Rh doping, also governs the two phase competition in this alloy system: the Rh 1\% system is chemically compressed near to the AF instability point, already at ambient pressure. The hysteretic behavior is also detected in the $\mu_o(\sigma)$ curve, supporting the argument that the transition is of first-order.

In our previous measurements using hydrostatic pressure, we observed a sudden increase in \moment from $\sim$ 0.22 to $\sim$ 0.40 \mb at $P_c \sim 1.5\ {\rm GPa}$. If this anomaly is also caused by the increase in $\hat{\eta}$, then similar behavior should be observed at $\sigma$ ($\bot$\,\dir{001}) $\sim 0.6\ {\rm GPa}$, where $\hat{\eta}$ is expected to reach the value ($\sim 1.8 \times 10^{-3}$) estimated at $P_c$. The maximum applied $\sigma$ in the present study is 0.61 GPa ($||$\, \dir{110}), and in this $\sigma$ range we observed no indication of the $P$-transition (see Fig.\ \ref{fig:moment-sc-uniaxial} and Fig.\ \ref{fig:scale}: upper right data points). Further investigation with higher stress will be needed to resolve the origin of this anomaly.

\subsection{The application of the Landau theory}
The stress-induced first-order phase transition observed in \urs is qualitatively understood in terms of the Landau's free energy theory with a time-reversal-invariant order parameter as follows. We assume the free energy $F(\psi,M)$ of the form, \cite{rf:Shah2000,rf:Chandra2001,rf:Liu73, rf:Motoyama2003}
\begin{eqnarray}
F&=&\frac{1}{2}r_\psi\psi^2+u_\psi\psi^4+\frac{1}{2}r_MM^2+u_MM^4\\ \nonumber
&&{}+2u_{\psi M}\psi^2M^2,\\
 r_\psi&=&a_\psi(T-T_\psi),\\
 r_M&=&a_M(T-T_M),
 \end{eqnarray}
where $\psi$ and $M$ denote the hidden order parameter and the staggered magnetization, and the signs of $a_i$ and $u_i$ are positive. It is straightforwardly seen that a first order phase transition between $\psi$ and $M$ may occur at the boundary $r_\psi=\sqrt{u_\psi/u_M}r_M\ (<0)$ on the condition $u_\psi u_M<u_{\psi M}^2$. Suppose that only the symmetry invariant strains are relevant. Then the total free energy, $F_{t}=F+F_{el}+F_{me}$, including the elastic energy $F_{el}$ and the magneto-elastic energy $F_{me}$ becomes,
\begin{eqnarray}
F_{t}&=&\frac{1}{2}r_\psi\psi^2+u_\psi\psi^4+\frac{1}{2}r_M'M^2+u_MM^4\nonumber \\
&&{}+2u_{\psi M}\psi^2M^2\nonumber\\
&&{}+\frac{1}{6}c^{\alpha 1}\hat{v}^2+\frac{\sqrt{2}}{3}c^{\alpha 12}\hat{v}\hat{\eta}+\frac{1}{3}c^{\alpha 2}\hat{\eta}^2, \label{eq:total-energy}\\
r_M'&=&a_M(T-(T_M+\frac{2D_v}{a_M}\hat{v}+\frac{2D_\eta}{a_M}\hat{\eta})).
\end{eqnarray}

\begin{figure}[tbp]
\begin{center}
\includegraphics[keepaspectratio,width=0.4\textwidth] {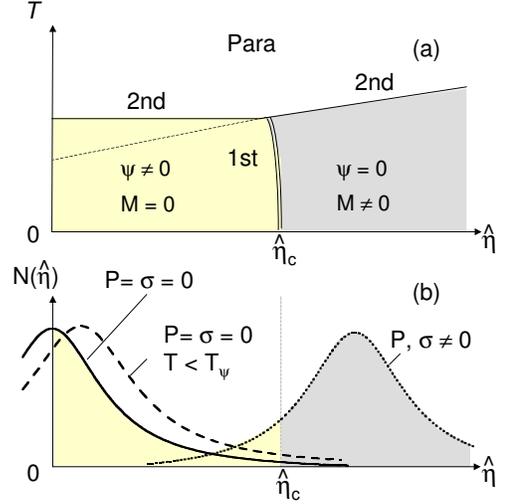}
\end{center}
	\caption{(a) The schematic drawing of the $\hat{\eta}-T$ phase diagram expected from the Landau's free energy consideration involving the elastic and magneto-elastic interactions. The phase boundary between hidden order and the AF order ($\hat{\eta}=\hat{\eta}_c$) is characterized by the first order phase transition. The $\hat{\eta}_c$ value of \urs is estimated to be $\sim$ 10$^{-3}$. (b) The schematic drawing of the distribution of $\hat{\eta}$. The width of distribution is expected to be $\sim$ 10$^{-4}$. The distribution is shifted right by thermal expansion caused by hidden order (the broken line) and by compression (the dotted line), generating the inhomogeneous AF phase. }
	\label{fig:landau-eta}
\end{figure}

Here, we neglected the coupling between $\psi$ and the strains for simplicity, but it should be remembered that $\psi$ seems to be also weakly coupled to $\eta$. This is expected because $\eta$ increases below \tord, \cite{rf:deVisser86} and because \tord increases with $P$ \cite{rf:Louis86, rf:McElfresh87, rf:Fisher90, rf:Ido93, rf:Oomi94} and $\sigma$($||$\,\dir{100}). \cite{rf:Bakker92} $T_\psi$ must be larger than $T_M$ at ambient pressure, since hidden order forms the majority phase. If $D_\eta>0$ and $|D_\eta|\gg|D_v|$ in this situation, then the AF transition temperature $T_M'$ ($\equiv T_M+\frac{2D_v}{a_M}\hat{v}+\frac{2D_\eta}{a_M}\hat{\eta}$) increases with increasing $\hat{\eta}$, so that the first order phase transition occurs at the critical point $\hat{\eta}_c$ as shown in Fig.\ \ref{fig:landau-eta}(a). By comparing the expected phase diagram with the present experimental results, $\hat{\eta}_c$ is roughly estimated to be $\sim 10^{-3}$ in \urs . Since $\hat{\eta}$ is an extensive variable, in principle the phase diagram should have an area near $\hat{\eta}_{c}$ where $\hat{\eta}$ shows a discontinuous change. Such an area is, however, expected to be very narrow, \cite{rf:Motoyama2002, rf:Motoyama2003} and not described in Fig. 7.

The above consideration is intended for a homogeneous system, and does not account for the inhomogeneous development of the AF phase. The crucial feature would be the smallness of $\hat{\eta}_c$. Here we suggest the presence of random distribution of $\eta$ in \urs , due to some imperfection of the crystal, as schematically shown in Fig.\ \ref{fig:landau-eta}(b). The width of the distribution is expected to be of the order of $10^{-4}$, which will be hard to detect and analyze using the usual microscopic probes. At ambient pressure, the mean value of $\hat{\eta}$ (taken as 0 in Fig.\ \ref{fig:landau-eta}(b)) should be smaller than $\hat{\eta}_c$, so that most part of the sample shows hidden order below $T_{\psi}$. We should remember here that the linear thermal-expansion coefficients show an increase of $\hat{\eta}$ of the order of 10$^{-4}$ below $T_{\psi}$. \cite{rf:deVisser86} $\hat{\eta}$ is thus expected to exceed $\hat{\eta}_c$ in small fragmentary regions of the sample, where the AF order takes place, being detected as tiny moment on volume-average. By applying $P$ or $\sigma\,\bot\,[001]$, the mean value of $\hat{\eta}$ exceeds $\hat{\eta}_c$, and the AF volume fraction inhomogeneously develops to the whole part of the sample, as is observed in the $^{29}$Si-NMR measurements under $P$. \cite{rf:Matsuda2001,rf:Matsuda2003} The temperature and stress dependence of the AF volume fraction should strongly depend on the condition of sample preparation, because such has a strong influence on the compressibility, the thermal expansion, and the distribution function of $\hat{\eta}$. This is consistent with the observed annealing effects, where the magnitude, the onset temperature and the $T$ variation of the AF Bragg-peak intensity all show significant sample-quality dependence. \cite{rf:Fak96} In particular, in this context the onset temperature of $I(T)$, which we define as $T_m^+$ in this paper, could become higher than \tord , if the distribution of $\hat{\eta}$ extends over $\hat{\eta}_c$ above \tord . This is actually observed in the present system, \cite{rf:Broholm87,rf:Fak96,rf:Ami99,rf:Isaacs90,rf:Honma99} where the width of onset $|T_m^+ - T_m^- |$ strongly depends on the specific experiment and sample. We emphasize that the AF response of such variety of starting conditions at ambient pressure is dominated by undetectably small change in the $c/a$ ratio.

Through the above considerations, we have stressed that the weak magnetism at ambient (and very low) pressure could reasonably be understood as the mixing of the high-pressure AF phase. This allows ones to adopt a time-reversal-invariant hidden-order parameter such as quadrupole moment. However, the presence of the AF fraction at very low pressure has not yet been confirmed by experiments. The present experiments do not exclude the possibility that the low-pressure magnetism is induced by an order parameter that breaks time reversal invariance but is nearly non-magnetic, such as an octupole moment. \cite{rf:Kiss2004}

\section{Conclusion}
We have presented elastic neutron scattering experiments under uniaxial stress on single crystal \urs , and discussed the nature of the unusual competition between hidden order and inhomogeneous AF order. A significant increase of the AF Bragg-peak intensity was observed when $\sigma$ is applied along the [100] and [110] axes, while it is nearly constant for {\sigmatoc}. The $\sigma$ variation of the AF scattering intensity for \sigmatoa roughly corresponds with that for {\sigmatob}, indicating that the AF evolution is isotropic with respect to compression in the tetragonal basal plane. The isothermal curve of the AF Bragg-peak intensity, which was obtained for U(Ru$_{0.99}$Rh$_{0.01}$)$_2$Si$_2$ as well as \urs under the zero-stress-cooled condition, shows a clear hysteresis loop, indicating that the phase transition from hidden order to the AF order is of a first-order. It was also found that the application of uniaxial stress enlarges the AF phase more effectively than that of hydrostatic pressure. We considered the crystal distortions induced under $\sigma$ and $P$, and pointed out that the observed features can reasonably be explained by the increase of the $c/a$ ratio with the compression. This interpretation is consistent with the results of the recent thermal-expansion measurements performed under hydrostatic pressure. \cite{rf:Motoyama2003} The inhomogeneous development of the AF phase can also be ascribed to the presence of random axial strains with a very small distribution width of $\sim$ 10$^{-4}$. 

\begin{acknowledgments}
We are grateful to F.J.\ Ohkawa, T.\ Sakakibara, K.\ Nemoto and K.\ Kumagai for helpful discussions. This work was supported by a Grant-In-Aid for Scientific Research from Ministry of Education, Culture, Sport, Science and Technology of Japan. One of us (M.Y.) was supported by the Research Fellowship of the Japan Society for the Promotion of Science for Young Scientists.
\end{acknowledgments}


\end{document}